\RequirePackage[table]{xcolor}
\documentclass{anstrans}

\title{An Alternative to Stride-Based RNG for Monte Carlo Transport}
\author{Braxton S.~Cuneo,$^{*}$ Ilham Variansyah,$^{\dagger}$}

\institute{
$^{*}$CS Department, Seattle University, 901 12th Avenue, P.O. Box 222000
Seattle, WA 98122-1090, bcuneo@seattleu.edu
\and
$^{\dagger}$NSE Department, Oregon State University, 151 Batcheller Hall
1791 SW Campus Way 
Corvallis, OR 97331, variansi@oregonstate.edu
}

\usepackage{graphicx} 
\usepackage{booktabs} 
\usepackage{microtype} 
\usepackage{listings}

\usepackage{etoolbox}
\usepackage{xparse}

\begin{document}
\section{Introduction}

The techniques used to generate pseudo-random numbers for Monte Carlo (MC) applications bear many implications on the quality and speed of that program's work.
As a random number generator (RNG) slows, the production of random numbers begins to dominate runtime.
As RNG output grows in correlation, the final product becomes less reliable.

These difficulties are further compounded by the need for reproducibility and parallelism.
For reproducibility, the numbers generated to determine any outcome must be the same each time a simulation is run.
However, the concurrency that comes with most parallelism introduces race conditions. 
To have both reproducibility and concurrency, separate RNG states must be tracked for each independently schedulable unit of simulation, forming independent random number streams.

Historically, Monte Carlo radiation transport applications have relied upon efficient random number generators, such as Linear Congruential Generators (LCGs)~\cite{brown1994lcg}.
LCGs are very lightweight, tracking a state consisting of only one integer.

To support multiple ``independent'' random number streams in concurrent contexts, many MC transport applications make multiple instances of the same LCG but skip each instance's state ahead along the cycle of its pseudo-random sequence.
This seed-skipping approach works as long as one can reasonably cut the pseudo-random sequence of an LCG into segments, known as random number strides, that are large enough to accommodate the maximum possible per-instance sample count.

In MC transport, it is important to ensure that the size of the random number stride being used is sufficiently large for the applications of interest, especially those that are characterized by long particle event chains, as it has been reported that significant stride exceedance can cause bias in tally results~\cite{hakim2024flux}. Furthermore, careful treatment is needed to maintain reproducibility with asynchronous parallelism, domain decomposition, and different algorithm bases (e.g., using the event-based instead of the history-based algorithm for vector computers). While effective bookkeeping techniques exist for the generational eigenvalue transport calculations~\cite{brown1992reproducibility}, maintaining reproducibility in more general fixed-source or time-dependent problems is still very challenging.

We propose an alternative to the stride-based parallel LCG seeding approach that scales more practically with increased concurrency and workload by generating seeds through hashing and allowing for repeated outputs. Data gathered from normality tests of tally results from simple MC transport benchmark calculations indicates that the proposed hash-based RNG does not significantly affect the tally result normality property as compared to the conventional stride-based RNG.


\section{Linear Congruential Generator (LCG)}

The sequence produced by an LCG is defined by the following function, which is iteratively applied to an instance's current single-integer state, $X_n$, to produce its next state, $X_{n+1}$, which also serves as the output of the LCG:
\begin{equation}
    X_{n+1} = (g X_n + c ) \texttt{ mod } 2^m,
\end{equation}
where $g$ and $c$ are integer constants and $m$ is the number bits used.
Given that the integer used to store this state is finitely sized and that the output of the above iteration is bounded to modulo $2^m$, an LCG instance can only enter a limited number of states and will eventually repeat.
If parameters are chosen correctly, the cycle of an LCG can match $2^m$, meaning all numbers between 0 and $(2^m-1)$ exist within one cycle.

\section{Parallel LCG with Seed Striding}

Assuming that an LCG has a cycle length equal to $2^m$, any instances of that LCG can be located on its cyclic sequence.
The sequence of elements between the starting position of one instance and the starting position of the next instance  is the stride assigned to the earlier instance.
The strides in this sequence are essentially ``allocations'' within the cycle of pseudo-random values.
Any application that has multiple LCG instances must ensure that the stride allocated for an instance matches or exceeds the number of queries that instance may receive. Otherwise, some outputs may be repeated.

To perform such an allocation, an application must generate LCG states with guaranteed offsets relative to another state.
F. B. Brown provided an efficient way to do this through the following formula \cite{brown1994lcg}:
\begin{equation}
    \label{eq:skip-ahead}
    X_{k} = \prod_{j=0}^{m-1}\left(g^{2^j}\right)^{k_{(j)}}\texttt{mod} 2^m + c\left[1+g+...+g^{k-1}\right] \texttt{mod} 2^m.
\end{equation}
The second term of Eq.~(\ref{eq:skip-ahead}) may be evaluated in $O(m)$ steps rather than $O(k)$, allowing for quick calculation of an arbitrary number of LCG iterations $k$.

If the number of independently schedulable simulation units $u_0 ... u_{N-1}$ is some fixed $N$, and the maximum number of queries $Q$ is uniform across all units, the process of LCG sequence allocation reduces to assigning $X_{Qk}$ to $u_k$, where $X_0$ is the starting seed assigned to the first unit.
This strategy is analogous to a linear memory allocator, allocating contiguous sequences of strides in a pseudo-random cycle rather than allocating contiguous bytes in a buffer.
As long as the product of $Q$ and $N$ does not exceed $2^m$, this scheme should not produce any repeats across any outputs.
However, if this product does exceed $2^m$, the allocations will wrap fully around the LCG's cycle.

\subsection{Repetition and State Space Exhaustion}

Many applications have a dynamic number of independently scheduled units of simulation.
In Monte Carlo neutron transport, fission can increase the number of tracked particles, and capture can reduce that number.
Furthermore, some MC techniques promote particle splitting to reduce the variance of the quantities of interest.
It is not practical to predict when these events will occur or what they will effect.
Hence, the number of queries required by a simulation or its components can vary significantly.

To accommodate uncertainty in the number of queries, larger strides may be allocated for each unit, but this only works until $NQ$ surpasses $2^m$, and increasing workloads in modern Monte Carlo applications draw simulations closer to this limit. If a concurrent, deterministic application does not calibrate its allocations based on user input and does not rely upon linear allocation, it must speculatively allocate for the largest potential allocation requirements.

A natural strategy to address this is to relax the requirement for concurrency, speculatively allocating per-domain or per-source, and performing linear allocation locally.
While hierarchical stride allocation has been successful with domain decomposed problems \cite{brown}, it still cannot address the growing scale and concurrency of MC simulations.
For example, the ExaSMR project has simulated over $50$ billion particles per cycle over 40 cycles \cite{Frontier}.

Even if the scale of simulations ceased to grow, the continued development and adoption of GPU processing in large-scale MC applications makes the use of shared linear allocators more difficult.
Synchronizing execution between threads in different blocks carries significant overhead, which reduces the effectiveness of linear allocation.

Unless trends in simulation scale or GPU usage change, MC transport code developers essentially have three choices:
(1) use LCGs with larger states, (2) use an entirely different type of RNG that supports ``true'' separate RNG streams, or (3) allow applications to potentially re-visit LCG states.
In this work, we explore the third option, which is particularly attractive since it eliminates the need for stride allocation management without significantly increasing resource usage.

LCGs are so widely used not because they are the most random but because they are fast and have a low memory footprint.
Weaknesses in LCG randomness have been known for over a half-century \cite{planes}, yet LCGs are a staple of modern MC transport because they are simply ``random enough'' for the application.
Hence, LCG seeding schemes that maintain a comparably low correlation while allowing revisited states could be a boon for MC transport code developers.

\section{Parallel LCG with Seed Hashing}

A hash function is a function that can be used to map one or more values to some value of a fixed size.
The hash function used to collect data for this work is a modified version of \textbf{\texttt{murmur\_hash64a}}\footnote{https://github.com/aappleby/smhasher}, accepting one 64-bit value as a key and another 64-bit value as a seed.
This hash function works as shown in Listing \ref{lst:murmur}, with all values as 64-bit unsigned integers and with wrapping multiplications.
To frame the discussion, we will describe the proposed seed hashing techniques in terms of this function, hence referred to as $\texttt{hash}(k,s)$, with $k$ and $s$ acting as the key and seed inputs.
Nonetheless, these methods should be applicable to any randomized hash function.
\begin{lstlisting}[label=lst:murmur, caption=A C++ Murmur Hash with 64-bit keys/values.]
uint64_t  const M = 0xC6A4A7935BD1E995;
uint64_t subhash(uint64_t x) {
    return (x*M)^((x*M)>>47);
}
uint64_t hash(uint64_t k, uint64_t s) {
    utint64_t seed_hash = subhash(s^(8*M));
    return subhash(seed_hash^(subhash(k)*M));
}
\end{lstlisting}

\subsection{Seed Splitting}

In place of the linear allocation used in the seed striding approach per Eq.~(\ref{eq:skip-ahead}), a sequence of starting states, $S_1 ... S_N$, can be generated for a sequence of LCG instances by using the index of the instance as the key:
\begin{equation}
    S_k = \texttt{hash}(k,s), \quad k=1,2,...,N.
\end{equation}
If both the seed $s$ and the mapping of indexes to instance recipients are deterministic, the starting states assigned to each recipient must also be deterministic. This technique is useful in contexts where a linear allocator would be used, such as generating source particles for a particular fission or fixed-source batch. However, because no stride length is required to perform this calculation, users do not need to calibrate the stride length, and developers do not need to account for worst-case allocation scenarios.

\subsection{Hierarchical Seed Splitting}

The selection of the key for a linear allocation depends upon the context of the batch.
If different agents within the simulation need to perform the same batching process, but the batching should not be correlated, they should use seeds based on unique identifiers. For example, if a series of $I$ batches each need to create starting states for $J$ particles, one could chain multiple calls to the hash function, using the ID of the batch as the key for the first hash and the resulting value as the seed of the second hash:
\begin{equation}
    S_{i,j}= \texttt{hash}(j,S_i) = \texttt{hash}(j,\texttt{hash}(i,s)).
\end{equation}
It is better to use the batch ID for the first hash because it allows the first hash to be computed once for a given source and then supplied to the second hash that generates each particle's starting state. We call this process, where new seed states are generated from a parent state, seed hashing, or, more generally, seed splitting, which contrasts with the conventional approach of seed striding.

This seed-splitting method can be applied an arbitrary number of times for an arbitrary number of state-generating levels.
Given a single, global seed value, an application can produce a ``tree'' of deterministic seeds, with each seed generated through the seed of its parent context.
As long as parent identifiers are earlier in the hashing chain relative to child identifiers, as demonstrated above, only one hashing operation is required to calculate a seed if its parent seed is already known. 
Given this chaining method, each relevant ``phase'' of a simulation can be assigned an index used to create a per-phase global seed.
This obviates the need to track the number of previous seeds generated in a certain context by allowing the relevant entity to restart at a zero index with a new seed.

In cases where a full LCG is not required, such as for performing population control at the end of fission batches, seed hashing can be used to generate one-off deterministic pseudo-random values.
So long as each one-off value generated by a given parent seed uses a different key, the generated values should share little correlation.

\subsection{On-Demand Seed Splitting}

Finally, the seed-splitting technique can also be applied to cases where the number of seed recipients is not known ahead of time.
For example, it is generally unknown how often a particle would create secondary particles and how many secondaries are created at each event.
With the seed hashing technique, we can use the parent particle seed state $S_p$ in the hashing function to create the seed state of the secondary particle $S_c$:
\begin{equation}
    S_c = \texttt{hash}(k,S_p),
\end{equation}
where the order number of the secondary particle at the creation event can be used as the hash key $k$.
This on-demand seed-splitting technique does not require any coordination between other simulation units.
Hence, the application tracking these units does not need to synchronize their execution solely for the purpose of initializing newly created instances.
Furthermore, the methods used to store and transfer these units would not need to maintain ordering guarantees over said units.
To apply this in MC transport, all particles must include a random number state as part of their attributes. However, with this tradeoff, we achieve higher concurrency in secondary particle creation, with highly manageable reproducibility (even with domain decomposition and different MC algorithm basis).

\section{Implementation and Results}

The proposed hash-based RNG (LCG with hash-based seed splitting) has been implemented in the open-source Python-based Monte Carlo code MC/DC~\cite{variansyah2023mcdc} and tested across various problems and scales.
For this work, however, a Python script\footnote{https://github.com/CEMeNT-PSAAP/Parallel-RNG-Test} was written from scratch to run simple MC transport calculations by using either the hash-based RNG or the conventional stride-based RNG (LCG with seed striding).
The normality of the tally results and the calculation runtimes are compared to assess the relative applicability of the two RNGs.

While the calculations are run in serial, the RNG seeding is set up to achieve reproducibility in parallel simulations at the level of source particles and, for the hash-based RNG, secondary particles.
In the stride-based RNG, seed striding is performed for each source particle.
In the hash-based RNG, seed splitting is performed for each source and secondary particle.
The standard MCNP5 LCG parameters of Index 8~\cite{mcnp5rng} are used in both RNGs.

\begin{figure}[h]
    \centering
\includegraphics[width=0.48\textwidth]{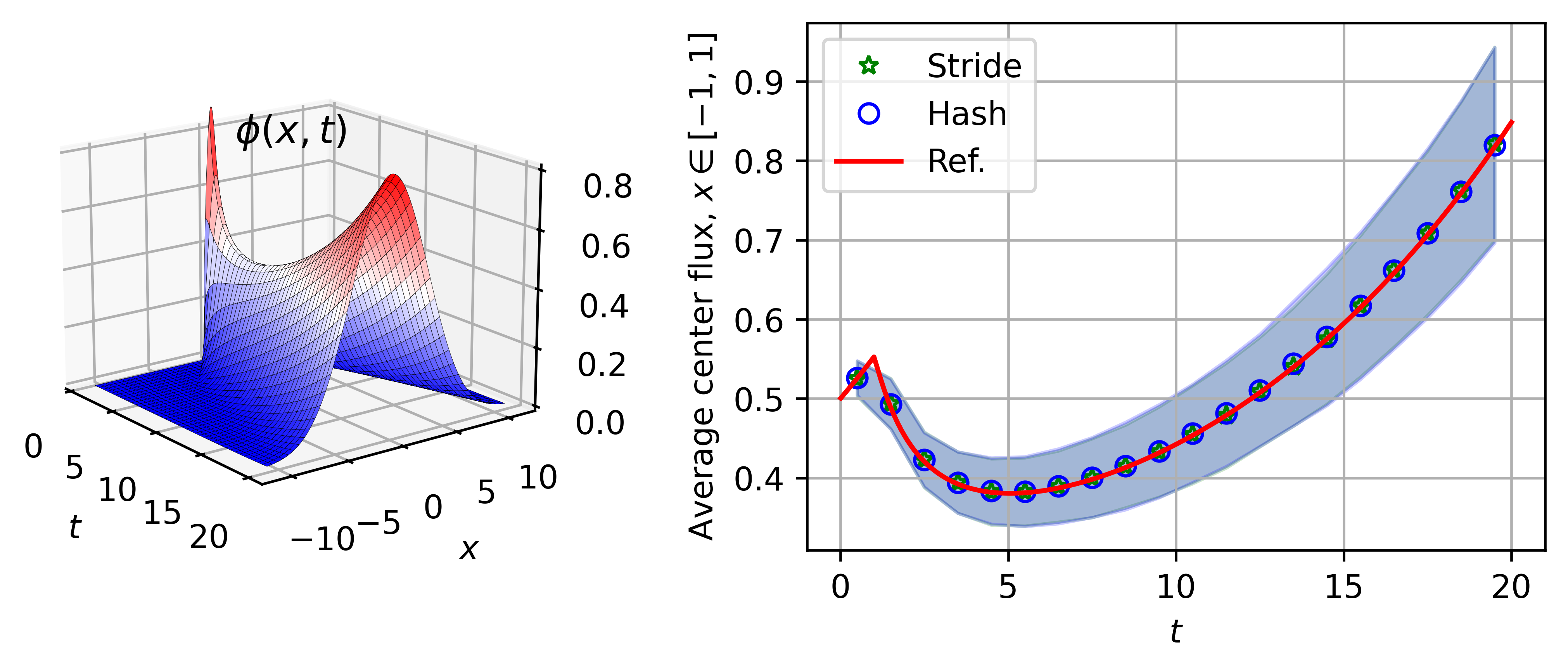}
    \caption{AZURV1 analytical flux solution $\phi(x,t)$ (left) and average center flux MC solutions calculated with the two RNGs (right). Overlapping shaded areas on the right figure represent the standard deviations of the 20 data points obtained from 1000 batch runs, each with 1000 source particles.}
    \label{fig:qoi}
\end{figure}

The test considers a supercritical time-dependent mono-energetic one-dimensional slab problem based on the benchmark AZURV1~\cite{Ganapol2001HomogeneousBenchmarks,variansyah2022pct}.
The particle transport is driven by a pulse at $t=x=0$ in the infinite homogeneous medium with isotropic scattering ratio $\Sigma_s/\Sigma_t=0.4$.
The scattering multiplicity is set to be 2.75, which adds up to the multiplication factor of 1.1.
The problem runs until $t=20$, at which point the population size grows by more than sevenfold.
The analytical particle flux solution $\phi(x,t)$ is shown in Fig.~\ref{fig:qoi}.
Quantities of interest tallied (with collision estimator) are average center fluxes in $x\in[-1,1]$ and unit-width time bins, which results in 20 data points also shown in Fig.~\ref{fig:qoi}.
The good agreement between the reference and the MC solutions calculated with stride- and hash-based RNGs verifies the accuracy of the implementations.

\begin{figure}[h]
    \centering
\includegraphics[width=0.48\textwidth]{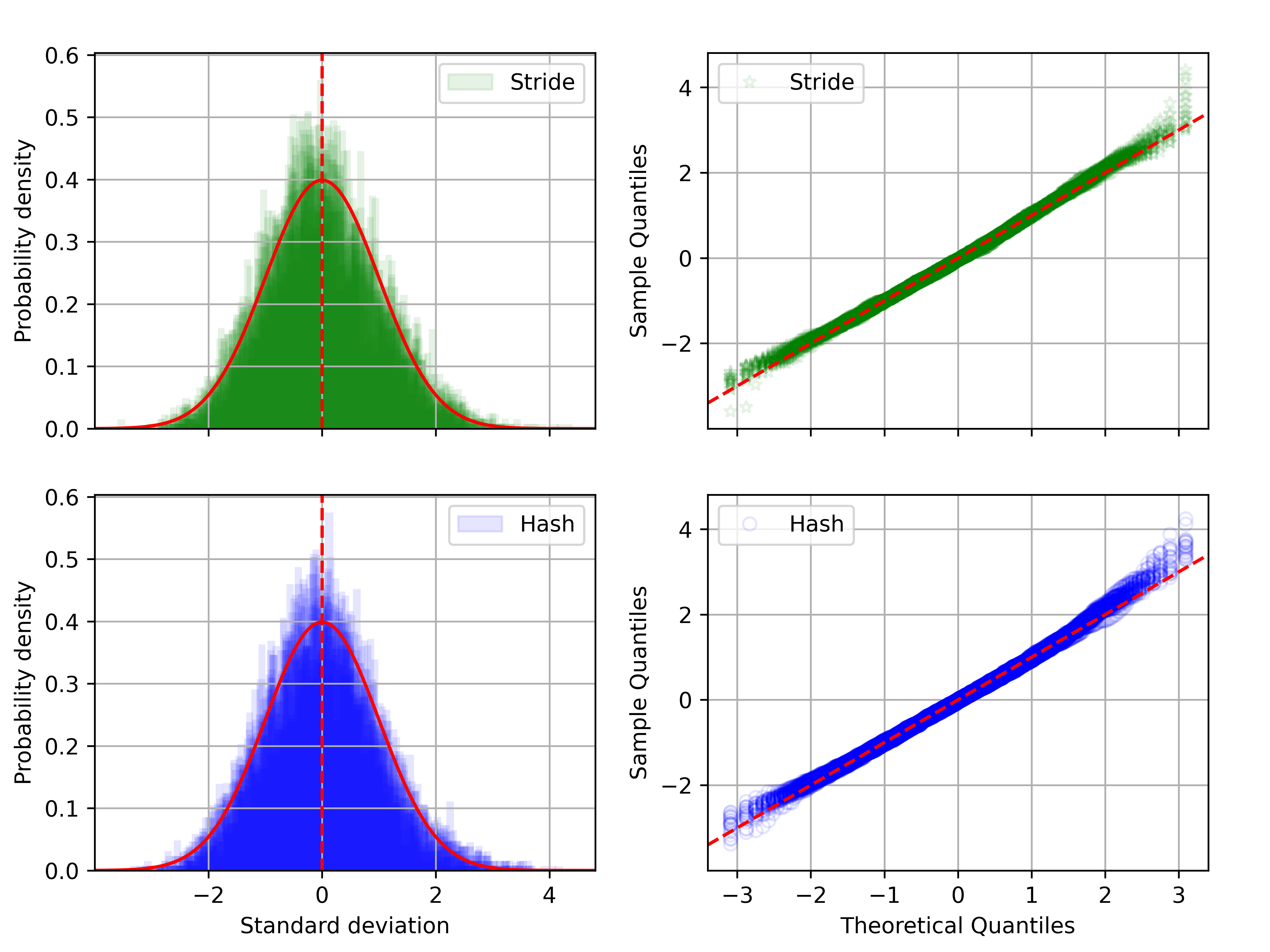}
    \caption{Overlapping distributions and quantile-quantile plots of the normalized deviations to the mean values of the 20 data points of the AZURV1 center flux calculated with the stride- and hash-based RNGs over 1000 batches.}
    \label{fig:normal_graph}
\end{figure}

The calculations are run in 1000 batches, each with 1000 source particles.
Normalized deviations to the mean values of the 20 data points over the 1000 batch runs are calculated, and their distributions are plotted (on top of each other) and compared to the expected normal distribution in Fig.~\ref{fig:normal_graph}.
Furthermore, to better assess the normality of the results, the associated quantile-quantile plots are also shown in Fig.~\ref{fig:normal_graph}.
It is found that both stride- and hash-based RNGs give tally solutions that appear to be normally distributed.

\begin{figure}[h]
    \centering
\includegraphics[width=0.4\textwidth]{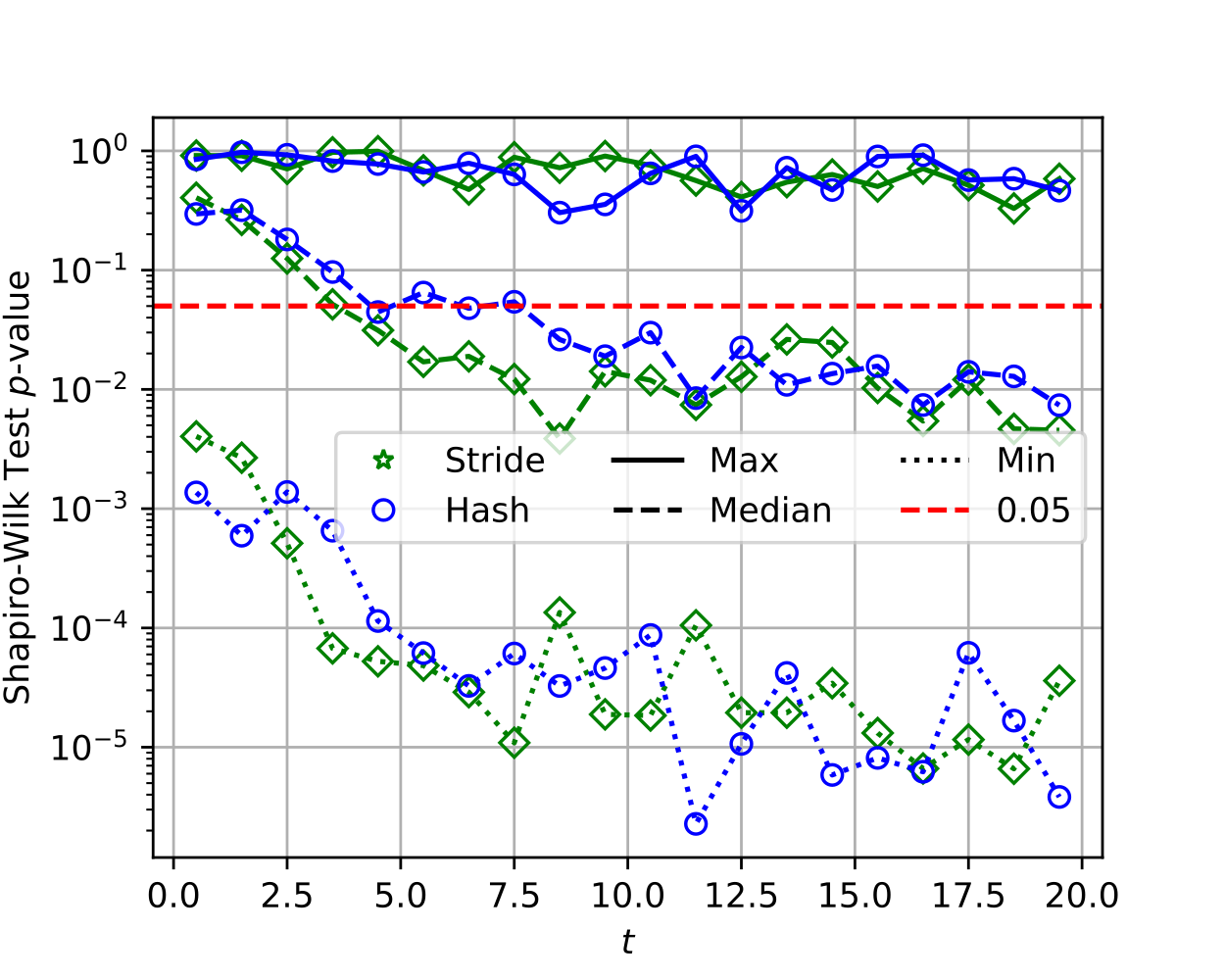}
    \caption{Saphiro-Wilk normality test results---maximum, minimum, and median $p$-values over 30 repetitions of the 1000 batch runs---at the 20 transient data points of the AZURV1 center flux calculated with the stride- and hash-based RNGs.}
    \label{fig:normal_test}
\end{figure}

We resort to the Shapiro-Wilk test~\cite{shapiro1965analysis} to get a more quantitative judgment on the normality of the MC tally results from the two RNGs.
The test is applied to each of the 20 transient data points that are calculated over 1000 batches.
The test is repeated multiple times to capture trends in the test results.
Seed splitting or seed striding is performed at each repetition, similar to the one done at each batch.
Figure~\ref{fig:normal_test} shows the resulting median, minimum, and maximum $p$-values over 30 repetitions.
The normality test results do fluctuate, as shown by the minimum and maximum values.
However, both stride- and hash-based RNGs exhibit similar trends and values.
It is worth mentioning that the $p$-values tend to decrease in time $t$, which can be attributed to the decreasing precision of the simulation as shown by the increasing shaded area (mean standard deviation) in Fig.~\ref{fig:qoi}.
Following the normality test convention, any $p$-value lower than 0.05 is flagged as failing the test.
From the total 600 tests performed (30 repetitions by 20 data points), there are 369 (61.5\%) and 339 (56.5\%) failing tests for stride- and hash-based RNGs, respectively.

As for runtime, 30 repetitions of 1000 batch runs, each with 1000 source particles, take about 113.26 and 111.46 seconds for stride- and hash-based RNGs. Considering uncertainties in runtime measurement, these averages are practically indistinguishable.

\section{Conclusion}

We propose a hash-based seed splitting as an alternative to the conventional seed striding RNG to achieve reproducible higher concurrency for MC particle transport calculations. Based on the graphical and standard normality tests performed in this work, both the conventional stride- and the proposed hash-based RNGs produce tally result distributions with similar degrees of normality.
This indicates that both MC transport RNGs are equally acceptable in terms of producing normally distributed results and maintaining the independent and identically distributed properties of the random variables.

\section{Acknowledgments}
This work was supported by the Center for Exascale Monte-Carlo Neutron Transport (CEMeNT) a PSAAP-III project funded by the Department of Energy, grant number: DE-NA003967.

\bibliographystyle{ans}
\bibliography{bibliography}
\end{document}